\documentclass[a4paper,twoside]{article}
\usepackage{graphicx}
\usepackage{color}
\usepackage{subfig}
\usepackage{multirow}

\newcommand{\params}{\bar{\theta}}
\newcommand{\model}{\mathcal{M}}

\newcommand{\be}{\begin{equation}}
\newcommand{\ee}{\end{equation}}
\newcommand{\bea}{\begin{eqnarray}}
\newcommand{\eea}{\end{eqnarray}}

\begin{document}

\title{Bayesian Model
  Averaging in Astrophysics: A Review} 

\author{David
  Parkinson$^{1}\thanks{Email: d.parkinson@uq.edu.au}$ and Andrew
  R.~Liddle$^{2}$}
\date{
$^1$ School of Mathematics and Physics, University of Queensland,
Brisbane, QLD 4072, Australia\\ 
$^2$ Astronomy Centre, University of Sussex, Brighton BN1 9QH, 
United Kingdom  
}
%\twocolumn[
%\begin{changemargin}{.8cm}{.5cm}
%\begin{minipage}{.9\textwidth}
%\vspace{-1cm}
    \maketitle

    \begin{abstract}
%\small{\bf Abstract:}     
We review the use of Bayesian Model Averaging in astrophysics. We
first introduce the statistical basis of Bayesian Model Selection and
Model Averaging. We discuss methods to calculate the model-averaged
posteriors, including Markov Chain Monte Carlo (MCMC), nested
sampling, Population Monte Carlo, and Reversible Jump MCMC. We then
review some applications of Bayesian Model Averaging in astrophysics,
including measurements of the dark energy and primordial power
spectrum parameters in cosmology, cluster weak lensing and
Sunyaev--Zel'dovich effect data, estimating distances to Cepheids, and
classifying variable stars.

    \end{abstract}
    \subsection*{Keywords}
cosmology: methods: data analysis  -- methods: statistical

%\medskip
%\medskip
%\end{minipage}
%\end{changemargin}
%]
%\small

\section{Introduction}
\label{firstpage}

We are in an unparalleled time in modern astrophysics, a {\it belle
  epoque} where more datasets and theoretical models are available
than ever before, with the number increasing at an exponential
rate. This is driven by Moore's law, as new technology has allowed the
development of better detectors to make observations and more powerful
computers to produce simulations.

In the middle, acting as the interface between these two areas, is the
field of statistical analysis, and again the growth in available
processing power has had a remarkable impact. Before the last decade,
Bayesian statistics were rarely applied in astrophysics, due to the
computationally intensive nature of computing the relevant quantities
accurately. Now posterior samples and model selection statistics can
be computed fast enough that the space of possible theories can be
fully investigated. Comprehensive overviews of Bayesian methodology
are given by MacKay, Gregory, and Sivia \& Skilling
\cite{MacKay,gregory05,SivSkil}; for a collection of articles on such methods as applied to
cosmology see ref \cite{BMIC}.

The strength of Bayesian methods is the ability to assign a
probability value to a model directly, based on the parameter ranges
allowed and the data available.  Model selection should proceed first,
before parameter estimation, and only once the best model has been
found should the parameters be estimated. Often though, the data is
not strong enough to distinguish decisively between the models.  In
this case, computing the parameter constraints under the assumption of
an individual model underestimate the true uncertainty of those
parameters.

The solution to this problem is Bayesian Model Averaging, where the
uncertainty as to the correct model is folded into the calculation of
the parameter probabilities.  Each individual posterior, generated
under the assumption of a particular model, is weighted by the model
likelihood and then combined with the others to give the
model-independent posterior. In this paper we review Bayesian Model
Averaging and its previous applications in astrophysics and cosmology.

In our review we found that Bayesian Model Averaging is used in two
manners. 
%Firstly there is the `prior mode', 
Firstly there are those situtations where the likelihood is
well understood and it is only the nature of the prior models and
parameters that are varied and averaged over. This is typically the
case in cosmology, where the cosmic microwave background (CMB) and
distance rulers are well measured observables. However, these
observations indicate the existence of new physical processes, such as
inflation and dark energy, for which there are no well-established
physical mechanisms. Then there are the %`likelihood mode', 
cases where the
physical model being investigated is well understood, but the model of
data acquisition or interpretation is more ambiguous, leading to
nuisance models that need to be averaged over. This is the case
for galaxy clusters, Cepheid variable distances, and star
classification. Of course this kind of division can be somewhat arbitrary,
as the models considered as nuisance for one investigator maybe of great
interest for another.

In Section \ref{sec:ModSelModAv} we introduce the ideas of Bayesian
Model Selection and Model Averaging. In Section \ref{sec:Methods} we
discuss some of the methods used to compute the model-averaged
posterior distributions, while in Section \ref{sec:Apps} we review the
applications of Bayesian Model Averaging that exists in the
astrophysics literature. In Section \ref{sec:Conc} we conclude.

\section{Model selection and model averaging}
\label{sec:ModSelModAv}

Bayes' theorem states the relationship between models ($\model$),
parameters ($\params$), and data ($D$)  
\be
P(\params|D,\model) =
\frac{P(D|\params,\model)P(\params|\model)}{P(D|\model)} \,, 
\ee 
where $P(\params|\model)$ is the prior probability distribution of the
parameters (assuming some model $\model$), $P(D|\params,\model)$ is
the likelihood, and $P(\params|D,\model)$ is the posterior probability
distribution of the parameters. The prior is updated to the posterior
by the likelihood. The term $P(D|\model)$ represents the model
likelihood, and is called the {\it evidence}. In the case of
single-model inference (where only a single model or set of parameters
is considered), the evidence is simply a normalizing constant, set to
satisfy the condition that the posterior distribution sums to unity.

However, in most interesting cases in astrophysics, the correct model
is not known, and the evidence takes different values for different
models. We can use Bayes' theorem again, at one level above, to
calculate the posterior odds between different models and perform {\it
  model selection},
\be
\frac{P(\model_1|D)}{P(\model_2|D)} =
\frac{P(D|\model_1)}{P(D|\model_2)} \frac{P(\model_1)}{P(\model_2)}
\,. 
\label{eq:modsel}
\ee 
Here $\model_1$ and $\model_2$ are the different models under
consideration, $P(\model_i)$ gives the prior probability of each
model, and $P(\model_i|D)$ is the model posterior probability.  Thus
the model posterior probability is updated from the prior by the
evidence.  If the model priors are equal then the ratio of posteriors
is simply the ratio of evidences.  The ratio of evidences is commonly
known as the Bayes factor $B$ \cite{KassRaferty}, and an
interpretation scale was suggested by \cite{Jeffreys} (though some
authors have started to use different language to qualify the
different levels, e.g.\ \cite{Trotta}). Many papers have already been
written about the use of the evidence for cosmological model
selection, a collection of the earliest being 
\cite{Jaffe96,DLW,JohnNarlikar,Slosar,Marshall2003,SWB,Niarchou,BCK}. 
For reviews see refs \cite{LidMukPark,Trottarev,Lidrev}.

The logical procedure would be to first perform a model selection
analysis to find the best model. Having done so, we would then perform
parameter inference for the parameters of that single best
model. However it is possible, even likely, that no model will have
decisive evidence ($\ln B > 5$) over all competing models. If we want
to include this model uncertainty in the parameter posteriors, we can
instead produce a {\it model-averaged} posterior distribution \cite{hoeting}, 
where the individual posteriors from each model are
summed together, weighted by the model posterior values:
\be 
P(\params|D) =
\frac{\sum_k P(\params|D,\model_k)P(\model_k|D) } {\sum_k
  P(\model_k|D)} \,.  
\ee 
This model-averaged posterior encodes the uncertainty as to the
correct model.

\section{Methods}
\label{sec:Methods}

Bayesian Model Averaging requires the computation of two quantities
per model, the individual model posterior and the
evidence. Astrophysical data are complex, requiring (at the very
least) differential equations to be integrated over time and/or space
to make predictions that can be tested against observations. As such,
almost no likelihood function can be solved for analytically, and so
the posterior and evidence have to be estimated by sampling methods.

Since such Bayesian methods, especially model selection and model
averaging, have only recently become popular in their application in
astrophysics, there are quite a number of methods available in the
statistics literature that remain unconsidered in this review. For
overview of some other methods see, for example ref \cite{ChenShaoIbrahim}.

\subsection{Markov Chain Monte Carlo}

Markov Chain Monte Carlo is a sampling method whose objective is to
generate a chain of sampled parameter values that obeys Markovian
statistics, and has the distribution being sampled as its equilibrium
distribution.  Estimates of the underlying moments of the probability
distribution are simply estimates of the moments of the chain, for
example the mean $\mu$ and variance $\sigma^2$ given by
\bea
\mu  =  E(\theta) & = & \frac1n \sum_{i=1}^{n} \theta_i \,, \\
\sigma^2  =  E((\theta-\mu)^2) & = & \frac1n \sum_{i=1}^{n} (\theta_i
-\mu)^2 \,. 
\eea
Similarly, we could use the chain to estimate the evidence,
\be
P(D|M) = E \simeq \frac1n \sum_{i=1}^{n} L(\theta_i){\rm Pr}(\theta_i)
\,, 
\ee
where $L(\theta)$ is the likelihood, $P(D|\theta,M)$, and ${\rm
Pr}(\theta)$ is the prior, $P(\theta|D)$, but this does not provide 
a valid (either unbiased or converging) approximation.   

Since samples cannot be drawn directly from the distribution
initially, MCMC conducts a random walk through the available parameter
space, generating points and adding them to the chain based on a set
of rules. The most common form of MCMC used in astrophysics is the
Metropolis--Hastings algorithm \cite{Metropolis1953,Hastings1970}.

MCMC has been used to generate posterior samples for a wide variety of
problems in astrophysics, becoming prevalent in cosmology following
the creation of the {\tt CosmoMC} code by Lewis and Bridle \cite{CosmoMC}. MCMC
methods are very efficient at tracing uni-modal posterior probability
distributions without pronounced degeneracies in the vicinity of the
best-fit region.  However, the evidence calculation receives
non-negligible contributions from the tails of the distribution,
because even though the likelihoods there are small, this region
occupies a large volume of the prior probability space. To compute the
evidence accurately, the entire prior space has to be sampled, which
can take a very long time using simple MCMC methods.

%A complex approach for using MCMC chains to compute
%the evidence has been suggested by van Haasteren (2009), but it remains
%to be demonstrated how rigorous or extensible this method is.

\subsection{Nested sampling}

\begin{figure}
\centering\includegraphics[scale=0.8]{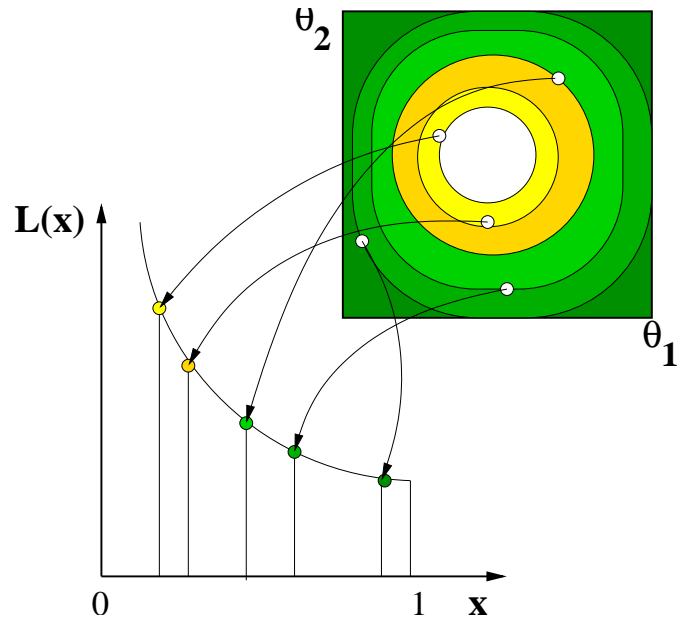}
\caption{A schematic of the nested sampling algorithm. The
  two-dimensional parameter space is shown at the top right. The
  points within it are considered to represent contours of constant
  likelihood, which sit within each other like layers of an onion
  (there is however no need for them to be simply connected). The
  volume corresponding to each thin shell of likelihood is computed by
  the algorithm, allowing the integral for the evidence to be
  accumulated as shown in the graph. (From \cite{Mukherjee2006})}
\label{fig:nestsamp}
\end{figure}

The nested sampling algorithm was devised by Skilling \cite{Skilling,Skill06} as
a method to compute the evidence. Nested sampling is a
Monte Carlo method that recasts the
multi-dimensional evidence integral as a one-dimensional integral
in terms of the prior mass X, where $dX = {\rm Pr}(\theta)d\theta$,
with $X$ running from 0 to 1. %The algorithm samples the prior
%recursively, assigning a `prior mass' to the thin shell of equal
%likelihood, then stripping away this shell and adding it to the
%integral in likelihood order. 
The integral is simply the sum of the likelihood-weighted prior mass samples,
\be
E = \int L(X)dX = \sum_{j=1}^m E_j~,~~E_j=\frac{L_j}{2}(X_{j-1}+X_{j+1})\,.
\ee
This is shown in Figure \ref{fig:nestsamp}.

In order to compute the integral accurately the prior mass is
logarithmically sampled. Firstly a set of $N$ `live' points is
randomly distributed within the prior parameter space. Then the lowest
likelihood point $L_j$ is discarded, replaced by a new point uniformly
sampled from the remaining prior mass (i.e.\ it must have a likelihood
greater than $L_j$). At every iteration the remaining prior mass,
$X_j$, shrinks by a known probabilistic factor that can be estimated
from the number of live points, and the evidence is incremented
accordingly. In this way the algorithm works its way towards the
higher likelihood regions.

Nested sampling was adapted for use in cosmology by Mukherjee,
{\it et al.} \cite{Mukherjee2006}, creating the {\tt CosmoNest} code, and by
\cite{MultiNest}, with {\tt MultiNest}. The key challenge
of applying nested sampling is the uniform sampling of the remaining prior
mass (as if the sampling is not uniform, the resulting Evidence value
will be biased). {\tt CosmoNest} achieved this by creating a single ellipsoid 
with contained all the remaining live points, and sampling uniformly from this shape. 
{\tt MultiNest} advanced this to automatically generate multiple ellipsoids, to cover multiple
likelihood peaks and increase the efficiency of the sampling process.

One of the advantages of using nested sampling is that the samples, as
well as contributing to the evidence integral, can also be used as
posterior samples. Although they don't obey Markovian statistics,
their posterior weight can easily be calculated from Bayes' theorem,
using their likelihood and the prior mass associated with them.

\subsection{Population Monte Carlo}

Population Monte Carlo is a generic adaptive importance sampling
method \cite{cappe:2004}. Importance sampling is the idea that you can
generate your samples completely stochastically from some
distribution. In importance sampling we draw samples from some
proposal distribution $q$, such that more samples are drawn from
regions which are judged to be {\it important}. The weights of each of
the samples is given as
\begin{equation}
w_i = \frac{P(X_i)}{q(X_i)} \,;~~~~ i=1,\ldots,N
\end{equation}
For example, if there is some new data that will update constraints
already in place from some previous experiment, you could simply
recompute the new posterior ($P$) at points given by a previous MCMC
chain (the values of $q$), importance sampling using the chain as the
proposal distribution. Obviously the distribution of points is
unchanged, but the new likelihoods give the new weights (or
importances) of the samples.

Similarly to MCMC, we can use importance sampling to compute the
evidence, since it will simply be the average of the weights,
\begin{equation}
E \approx \frac1N \sum_{i=1}^N w_i \,.
\end{equation}

Adaptive Importance Sampling (otherwise known as Population Monte
Carlo) updates the proposal distribution with each iteration,
attempting to move it closer to the distribution being sampled.  Douc
et al \cite{DGMR:2007} proposed to minimize the distance between the target
distribution $\pi$ (our posterior) and the proposal distribution $q$
as defined by the Kullback--Leibler divergence (the relative Shannon
entropy between the two distributions \cite{KullbackLeibler}). This
is defined to be
\begin{equation}
D_{\rm KL}(p||q) = \int p(\theta) \log\frac{p(\theta)}{q(\theta)} d\theta \,,
\label{KLdiv}
\end{equation}
where $p$ and $q$ are two distributions, and $x$ is a set of variables that
the distributions depend on. Rather than integrate over the parameter
space, we can compute this function by simply summing it over the
samples, i.e.
\begin{equation}
D_{\rm KL}(\pi||q) \approx \sum_{i=1}^N P(\theta_i) \log
\frac{P(\theta_i)}{q(\theta_i)} \,. 
\end{equation}

Kilbinger et al.\ \cite{Kilbinger:2009}, following the approach of Capp{\'e} et
al.\ \cite{cappe:2004,cappe:2008}, applied Population Monte-Carlo to cosmology, using
an implementation where the importance function $q$ were modelled by a
sum of mixtures
\be
q_X(x) = \sum_{d=1}^D \alpha_d f_{Y_d}(x) \,.
\ee
where $\alpha=(\alpha_1,\dots,\alpha_D)$ is a vector of adaptable
weights for the mixture components (with $\alpha_d > 0$ and $
\sum_{d=1}^D \alpha_d=1$) and $Y_d$ a vector that specifies the
components. Here $f$ is a parameterized probability density function,
usually taken to be a multi-variate Gaussian or Student-t
distribution. In these cases, the updated weights and distribution
parameters that minimise the KL divergence between the two
distributions can be solved for analytically, details of which can be
found in ref. \cite{cappe:2008}.

\begin{table*}

\centering
\caption{\label{tab:cosmodates}List of cosmological parameters and
  their approximate date of introduction into the standard
  cosmological model.}
\label{standardparameters}
\resizebox{\textwidth}{!}{%
\begin{tabular}{lccl} \hline
Cosmological parameter & Symbol & Introduction date& Alternate
versions \\ \hline 
Hubble parameter & $H_0$ & 1929 & Distance to last scattering surface
$\Theta$ \\ 
Matter density & $\Omega_{\rm m}$ & 1924 & Physical matter density,
$\omega_{\rm m}=\Omega_{\rm m} h^2$ \\ 
CMB Temperature & $T_{\rm CMB}$ & 1964 & Radiation density
$\Omega_{\rm r}$ \\
Cold Dark Matter density & $\Omega_{\rm c}$ & (1933) 1975 & \\
Galaxy Bias & $b$ & 1985 & \\
Amplitude of density perturbations & $A_{\rm s}$ &  1992 &CMB
normalization $\Delta T/T$, dispersion at $8h^{-1}$Mpc, $\sigma_8$
\\  
Optical depth to reionization & $\tau$ & 2003 &  Redshift of
reionization, $z_{\rm reion}$ \\ 
Spectral index of scalar perturbations & $n_{\rm s}$ & 2006 \\ \hline
\end{tabular}}
\end{table*}

Population Monte Carlo suffers from some of the problems faced by
ordinary MCMC. For example, when the mixture distribution doesn't
cover all the modes in a multi-modal distribution, it will fail to
pick up contributions from these modes.

One advantage of PMC is that it allows us to break somewhat from the
`sequential tyranny' of ordinary Markov Chain methods. Since the
importance function is known, large numbers of samples can be
generated from it {\em in parallel}, making it ideally suited for use
in Graphics Processing Unit (GPU) computing, since the GPUs can
perform a very large number of parallel floating point operations. Of
course, the update step still requires these samples to be brought
together and analysed so that the importance function can be
updated. Kilbinger et al.\ \cite{Kilbinger:2009} state that PMC gives an unbiased
estimator of the evidence at each iteration, and will converge on a high
accuracy value in a very small number of such iterations.

As with Nested Sampling, PMC produces samples of the posterior, and
so can be used for generating model averaged posteriors.

\subsection{Transdimensional MCMC}

The biggest time requirement in using these previous methods for
Bayesian Model Averaging is that the posteriors and evidences for each
model need to be computed for each model individually. It could be
more efficient to instead do all of this in one step, simultaneously sampling
from the model space as well as the parameter values. For this we need
an MCMC algorithm that can change dimensionality, while still
maintaining reversibility to guarantee the chains elements satisfy
detailed balance. Examples of transdimensional Monte Carlo include
Reversible Jump MCMC \cite{RJMCMC} birth-and-death process 
Monte-Carlo \cite{Stephens2000}, and Continuous Time MCMC \cite{CRR}.

Reversible Jump MCMC is an algorithm in which the dimensionality of
the model can be changed during the chain, by allowing transitions
between models as well as parameter values. In order to maintain
detailed balance, Green (1995) proposed a new acceptance probability.
The general acceptance ratio from state 
$X =(1,\theta^{(1)})$ to state  $X =(2,\theta^{(2)})$, where 1 and 2
represent the model spaces that the two different states occupy, as
\begin{eqnarray}
 \alpha_m (X,X') & =&  \\ \nonumber 
&& \hspace*{-2cm}  {\rm min} \left\{ 1,
 \frac{P(2,\theta^{(2)}|D)j(2,\theta^{(2)})q_2(u^{(2)})}{P(1,\theta^{(1)}|D)j(1,\theta^{(1)})q_1(u^{(1)})}
 \left\| \frac{\partial
   g(\theta^{(2)},u^{(2)})}{\partial(\theta^{(1)},u^{(1)})} \right\|
 \right\} \,,  
\end{eqnarray}
where $P(\theta|D)$ is the usual posterior value, $q$ is the
probability density for parameter values inside that model, $j $ is
the probability for steps between models, and $u$ is an extra random
variable that controls the proposal of different models.  The final
term is a Jacobian of the function $g(\theta,u)$ that converts
parameter values between models, i.e. $(\theta^{(1)},u^{(1)}) =
g(\theta^{(2)},u^{(2)})$. If these are independent, then the Jacobian
is simply unity.

The drawback of RJMCMC is that in practice, for the case of a small
number of models, it isn't more efficient to sample simultaneously in
both model and parameter space, since the steps between models often
have a low acceptance rate. Instead, RJMCMC is more useful in cases
where the dimensionality of the problem is large.  For example, it has
been applied in astrophysics to detection problems where the number of
sources is unknown, such as in gravitational wave astronomy \cite{CorLitt}.

\section{Applications in Astrophysics}
\label{sec:Apps}

\subsection{Cosmological Parameters}

Cosmological model development has proceeded incrementally, with a new
parameter being added to the concordance cosmological model at a rate
of approximately one a decade (though starting to increase faster in
recent years: see Table \ref{tab:cosmodates} for an approximate
history). Cosmological models rarely make specific predictions for the
values of the parameters.  For example, there is no theory that
predicts exactly the value of the Hubble parameter, though you can
infer a plausible range from other data, such as the age of the oldest
stars in the Universe. The models that do predict values for the
observables, for example models of inflation that predict specific
values of $n_{\rm s}$ (the spectral index) and $r$ (the
tensor-to-scalar amplitude ratio), do so by changing the free
parameter to something more fundamental to the theory (e.g.\ the value
of the mass scale or coupling in the inflationary potential).

Therefore, because there are no fundamental models that make very
accurate predictions for the values of the cosmological parameters,
they are phenomenological parameters. They can be measured accurately,
but little or no fundamental physics can (yet) be derived from them.
This isn't true for all cosmological parameters, and may not be always
true for others. The optical depth to reionization, for example, could
be accurately predicted if we had a perfect model for the formation of
the first generation of stars and the reionization of the cosmic
hydrogen. Alas, perfect models are found only in mathematics, and
physicists have to live in the real Universe.

The lack of a good theory to predict these parameters means that
the relevant priors may be subject to a level of contention.

\subsubsection{Dark Energy}
\label{sec:DEonly}

The mysterious dark energy makes up around 70\% of the energy density
of the Universe, and acts against gravity to drive an accelerating
expansion. Its discovery in the late 1990s was largely unexpected, and
even today there are no very compelling models for where it comes
from.  The Dark Energy Task force report stated that ``{\it the nature
  of dark energy ranks among the very most compelling of all
  outstanding problems in physical science.}''  \cite{DETFreport}.

The dynamical properties of the dark energy are normally parameterized
in terms of the equation of state parameter $w$, where
\be
w = \frac{P}{\rho} \,.
\ee
Here $P$ is the pressure of the dark energy fluid and $\rho$ is the
density.  Different models of the dark energy make different
predictions for the value of the equation of state. The cosmological
constant model, where the dark energy is the zero-point energy of the
vacuum, predicts a value of $w=-1$ only. Quintessence models, where
canonical scalar fields are slowly rolling down some potential, can
take time-dependent values in the range $-1 < w <1$.

\begin{figure}
\centering
\includegraphics[angle=-90,scale=0.9]{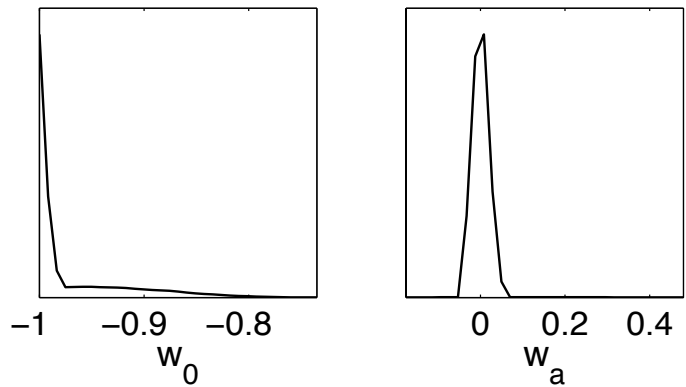}
\caption{Posterior parameter distributions for dark energy equation of
  state parameters $w_0$ (the value today) and $w_a$ (the rate of
  change with scale factor). These distributions were obtained using
  the WMAP+SDSS(BAO)+SNLS data combination averaging over the
  quintessence-type models (I, II and V) alone. Some smoothing of the
  delta-functions has been carried out by binning. (From \cite{Liddle:2006kn})}
\label{fig:bma_de}
\end{figure}

Liddle et al.\ \cite{Liddle:2006kn} addressed the problem of model
independent constraints on the equation of state parameter $w$ where
the correct model is not known. They used five different models for
the equation of state, which were:
\begin{enumerate}
\item The cosmological constant model, where $w=-1$ at all times,
\item A constant equation of state model, where $w$ can vary in the
  range $-1 < w < -1/3$, 
\item A constant equation of state model, where $w$ can vary in the
  wider range $-2 < w < -1/3$, 
\item A time-varying equation of state, where the two parameters have
  prior range $-2 < w_0 < -1/3$ and $ -4/3 < w_a < 4/3$, 
\item A time varying equation of state, where the equation of state as
  a function of scale factor has prior range $-1 < w(a) < 1$. 
\end{enumerate}
Models (ii) and (v) placed priors on the equation of state that
excluded some of the high-likelihood region found in models (iii) and
(iv), but these models were created to enforce the weak energy
condition.  In models (iv) and (v) the CPL parameterization
\cite{Linder2003,ChevPol} was used, where \be w(a) =
w_0 + (1-a)w_a \,.  \ee They also varied two parameters common between
all five models, the present fractional energy density of matter
$\Omega_{\rm m}$ and the value of the Hubble parameter today
$H_0$. They assumed a flat universe where the density of the dark
energy $\Omega_{\rm DE} = 1 - \Omega_{\rm m}$.

In terms of data used to compute the evidence of each of the models
and the posteriors of the parameters, they used distance data from the
then-current three-year Wilkinson Microwave Anisotropy Probe (WMAP)
observations \cite{Hinshaw07,Spergel07}, 
%in terms of the shift parameter $R$, 
the BAO measurement from the Sloan Digital
Sky Survey (SDSS) LRG sample \cite{Tegmark06}, and SN Ia data
from the HST/GOODS programme \cite{Riess04} and the first-year
Supernova Legacy Survey (SNLS \cite{Astier05}), together with nearby SN
Ia data. They computed the evidence and the posteriors simultaneously
using nested sampling.

The model selection results were inconclusive with no model favoured
and only model (v) significantly disfavoured. Of course, owing to the
nature of the different priors allowing different high likelihood
regions, the posteriors for the five models were very different. The
authors used Bayesian Model Averaging to plot the model-averaged
posterior for $w_0$ and $w_a$, averaging over all five models and also
those that enforced the weak energy condition (i, ii and v). The
results are shown in Figure \ref{fig:bma_de}.

\subsubsection{Curvature}

In \cite{Vardanyan2011}, the authors conducted a similar analysis
to Liddle {\it et al.} \cite{Liddle:2006kn}, which was described in section
\ref{sec:DEonly}, using more up-to-date data and including models for
the curvature of the universe simultaneously. They utilized data from
the WMAP five-year release \cite{Dunkley09}, once again using only
the distance data from the CMB, the updated BAO measurements from 
\cite{Percival07}, and the UNION08 SN-Ia data compilation \cite{Kowalski09}. 
They considered only three different models for the dark
energy, which were:
\begin{description}
\item{$\Lambda$:} The cosmological constant model, where $w=-1$ at all
  times, 
\item{W:} A constant equation of state model, where $w$ can vary in
  the range $-2 < w < -1/3$, and $w_a=0$, 
\item{WZ:} A time-varying equation of state, where the two parameters
  have prior range $-2 < w_0 < -1/3$ and $ -4/3 < w_a < 4/3$.
\end{description}

They also considered three models of the curvature of the universe:
flat, open, and closed. However, interestingly they considered these
three models with two different priors, effectively measuring two
different parameters. The Astronomers' prior considered $\Omega_k$ as
the parameter to be measured.  The Curvature scale prior considered
the curvature scale (the log of $\Omega_k$ or $-\Omega_k$ depending on
whether the model was open or closed) as the parameter to be
varied. They combined the three dark energy models and the three
curvature models together (creating nine different models in total)
for each of the two choices of prior.

\begin{figure}[tp]
\centering
\includegraphics[scale=0.35]{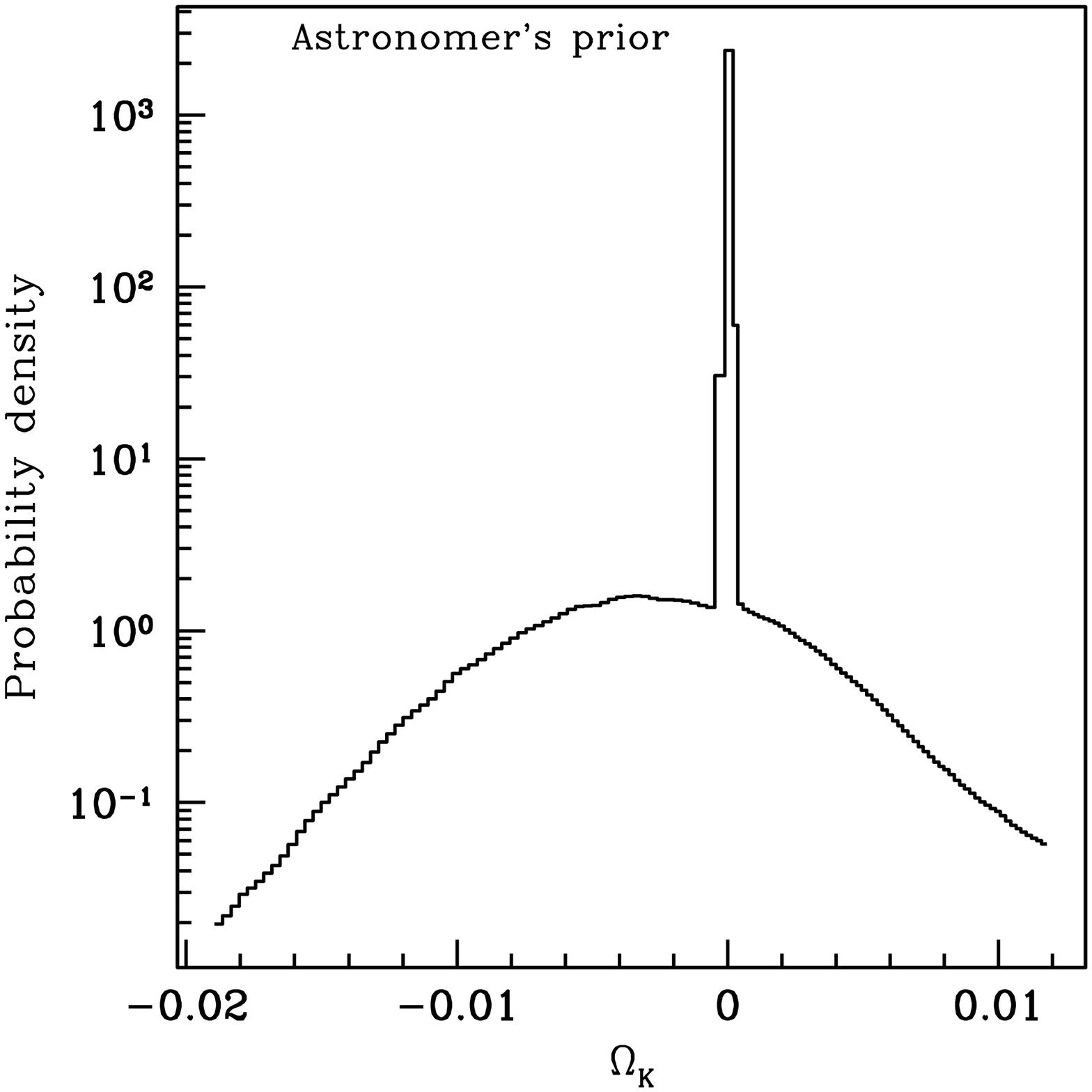}
\includegraphics[scale=0.35]{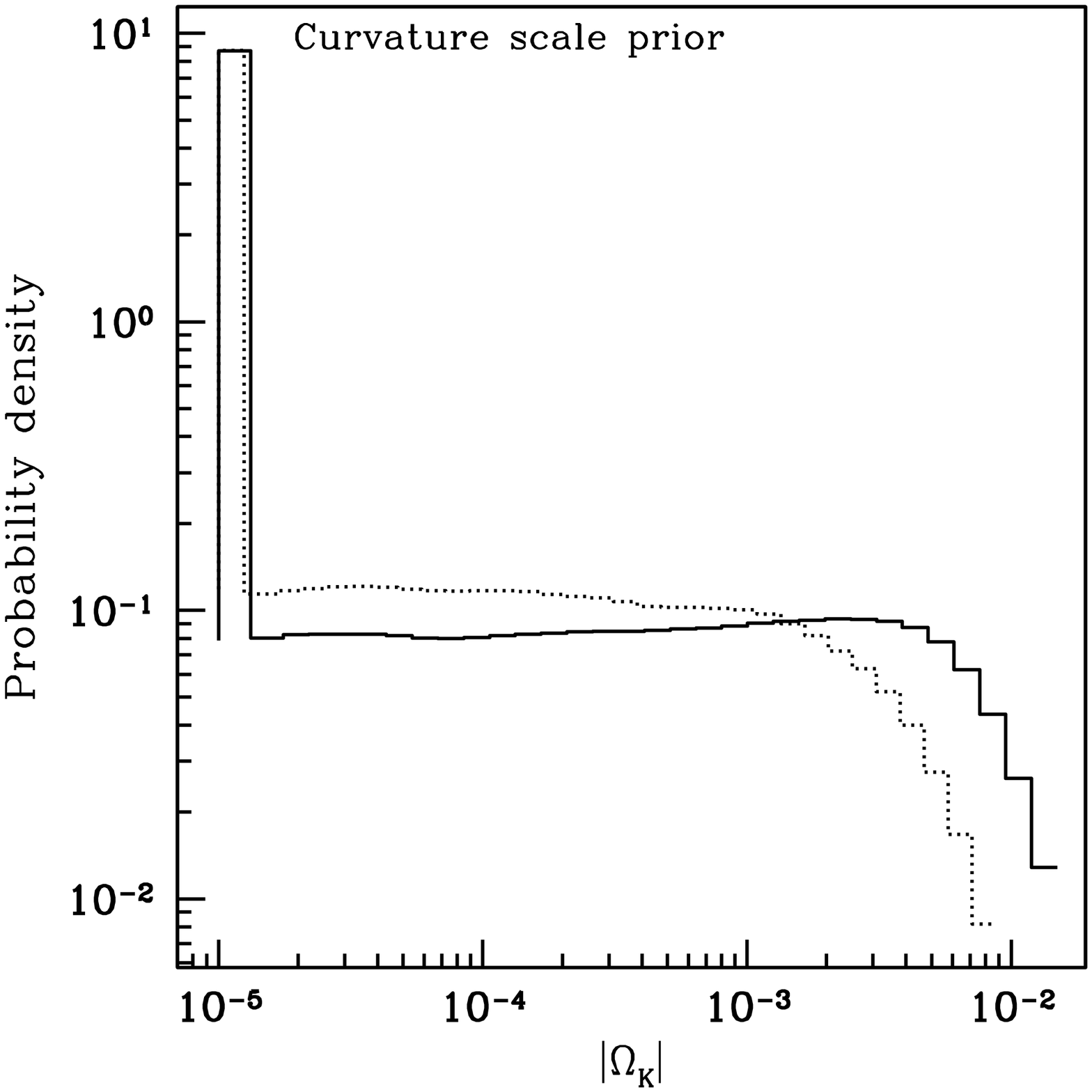}
\caption{Model-averaged posterior probability distribution for the
  curvature parameter, including all 9 models considered in the
  analysis, assuming the Astronomers' prior (top panel) and the
  Curvature scale prior (bottom panel) for the curvature parameter. In
  the bottom panel, the solid line applies to closed universes
  ($\Omega_k< 0$), while the dotted line to open universes ($\Omega_k
  < 0$). The peaks represent the Dirac delta function encompassing the
  probability mass associated with flat models. (From ref. \cite{Vardanyan2011})} 
\label{fig:bma_curv}
\end{figure}

They used Metropolis--Hastings Markov Chain Monte Carlo to compute the
posterior distributions and Bayes factors. They found that no model
was favoured over the flat $\Lambda$ model, but the closed or open
models with the Astronomers' prior were significantly disfavoured
compared to those same models using the Curvature scale prior. They
also produced model-averaging results for $\Omega_k$, averaged over
all nine models, but considering only either the Astronomers' prior or
the Curvature scale prior. The model-averaged posteriors are shown in
Figure \ref{fig:bma_curv}.

\subsubsection{Primordial Power Spectrum}

The models for the generation of the primordial spectrum of
fluctuations are rather better motivated than those of the dark
energy. Since the Big Bang model became established, many theories have
been proposed for the formation of structure in the Universe and
subsequently ruled out by observational evidence (e.g.\ active seed
models such as cosmic strings).

The primordial power spectrum of scalar perturbations is normally
modelled through a modified power-law function of wavenumber $k$, 
\be
\Delta^2_{\mathcal{R}}(k) =
\Delta_{\mathcal{R}}^2(k_*)\left(\frac{k}{k_*}\right)^{(n_{\rm s}-1)+\frac12
  \ln(k/k_*)n_{\rm run}} \,, 
\ee
where the amplitude is defined at a pivot scale ($k_*$), $n_{\rm s}$
is the spectral index (also known as the tilt), and $n_{\rm run}$ is
the running of the spectral index.

Harrison and Zel'dovich \cite{Harrison1970,Zeldovich1972} both suggested a
maximally-symmetric model, with equal power on all scales. In such a
model the spectral index is unity and the running is zero. Later, when
Guth \cite{Guth81} invented the concept of cosmological inflation, it was
realized that such a period of accelerated expansion would generate a
spectrum of density fluctuations with a spectral index close to unity
and a negligible running.

\begin{figure}
\centering
\includegraphics[scale=0.8]{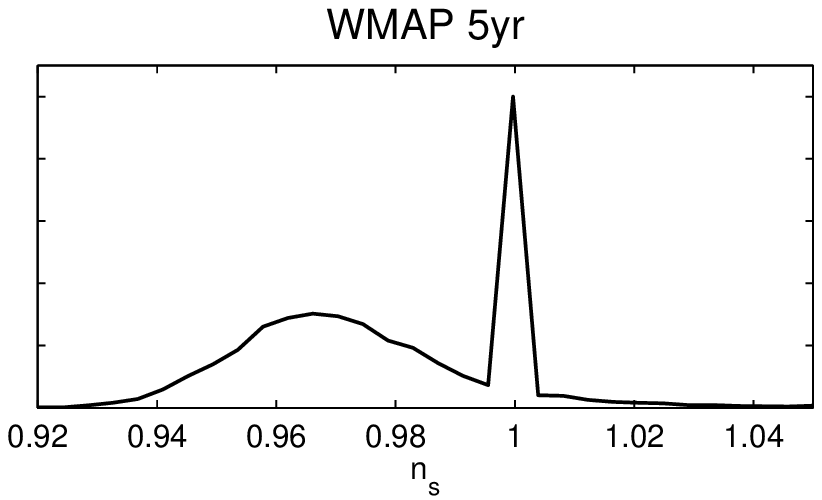}\\
\includegraphics[scale=0.8]{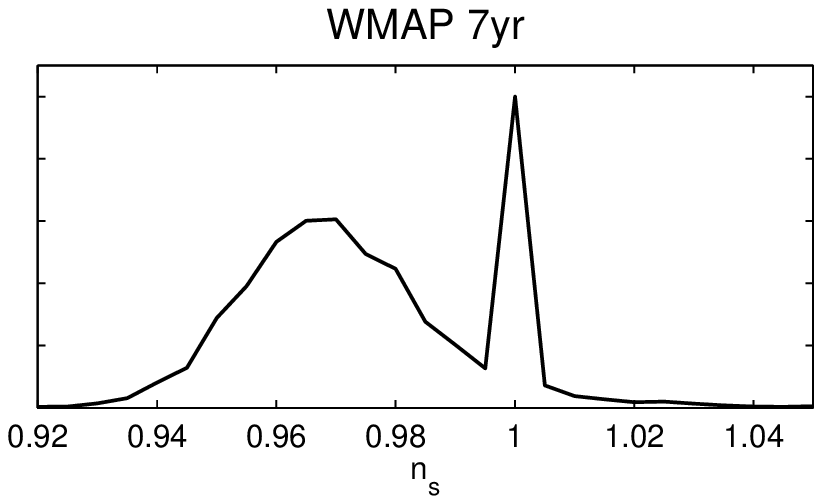}\\
\includegraphics[scale=0.8]{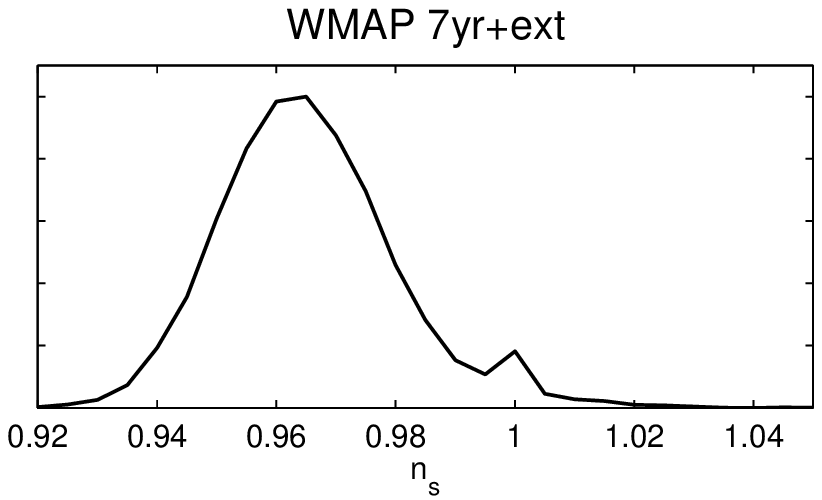}
\caption{The model-averaged posterior distributions for the spectral
  index $n_{\rm s}$, using the WMAP 5yr data only (top), the WMAP 7yr
  data only (middle) and the WMAP 7yr+ext compilation (bottom). The
  probability distribution includes a delta function around $n_{\rm
    s}=1$, artificially broadened in the plot by the
  binning process. (From ref. \cite{Parkinson:2010zr}) }
\label{fig:ns_averaged}
\end{figure}

Parkinson \& Liddle \cite{Parkinson:2010zr} applied Bayesian model averaging to
measurements of the parameters of the primordial power spectrum of
density fluctuations. They considered five different models.
\begin{enumerate}
\item A scale-invariant Harrison--Zel'dovich (HZ) spectrum of scalar
  perturbations  with no tensor component ($n_{\rm s}=1$, $r=0$). 
\item A tilted model, where the spectral index is allowed to vary,
  still with no tensors. 
\item A running model, where both the spectral index and the running
  of the spectrum ($n_{\rm run}$) are allowed to vary. 
\item A tensor model, where the spectral index of the scalar
  perturbations and the tensor-to-scalar amplitude ratio ($r$) are
  allowed to vary. 
\item A tensor+running model, where the spectral index,
  tensor-to-scalar ratio, and running all vary. 
\end{enumerate}

They also varied the physical baryon density $\Omega_{\rm b} h^2$, the
physical CDM density $\Omega_{\rm c} h^2$, the angular diameter of the
sound horizon at last scattering $\theta$, the optical depth to
reionization $\tau$, the (log) amplitude of the density fluctuations,
$\ln [10^{10}A_{\rm s}]$, and the amplitude of the Sunyaev--Zel'dovich
contribution to the CMB power spectrum $A_{\rm SZ}$. They used data
from the five-year and seven-year WMAP data releases, to compare how
much improvement in model selection occurred between them, and also
combined the seven-year WMAP with smaller-scale CMB experiments and
measurements of the galaxy power spectrum by the Sloan Digital Sky
Survey (SDSS).

The model-averaged posterior for $n_{\rm s}$ for the different data
compilations are shown in Figure \ref{fig:ns_averaged}. When WMAP data
alone was used, most of the models had similar evidence values, and so
the averaged posterior had a substantial contribution from the
delta-function from the HZ model. When the extra small scale and
galaxy power spectrum data were added, the models where $n_{\rm s}$
was allowed to vary had larger evidence values, and these models also
had posteriors for $n_{\rm s}$ that peaked for approximately the same
values.

\subsubsection{Summary}

An interesting point to note is that in each of the above cases
Bayesian model averaging was applied with a uniform model prior. That
is, none of the authors felt that one model could be favoured {\em a
  priori} over the others. Because of the mysterious nature of these
phenomena (dark energy, inflation) where any proposed physical
mechanism is still highly speculative, the authors were forced to
parameterize their ignorance with phenomenological descriptions. But
since it is unknown even if these descriptions are accurately
representing the real mechanism, no clear choice can be made
in advance. This meant that the dimension of the models played the
largest role in determining the evidence values, and so the
model-averaged posteriors were dominated by the models with the fewest
parameters. The exception to this was the primordial power spectrum
case, where the simplest model (HZ) is close to being ruled out, and
so its contributions to the model-averaged posterior are almost negligible.

\subsection{Cluster weak lensing and Sunyaev--Zel'dovich effect data}

The use of multiple models in a Bayesian framework is also useful in
the case where the nature of the data is uncertain. One such case is
galaxy clusters. Galaxy clusters are the largest gravitationally-bound
objects in the Universe, and as such can be used as cosmological
probes. However, the distribution of matter inside the cluster is
crucial in recovering measurements of their properties, and often some
model has to be assumed before probabilistic inferences on other
cosmological parameters can be made.

Marshall et al. \cite{Marshall2003} applied Bayesian methods to the
problem of cluster data modelling, attempting to fit the best model to
the clusters' gas and temperature profiles using simulated
Sunyaev--Zel'dovich effect data and weak gravitational lensing
data. They considered two models of the gas density profile, the
`Beta' model (see e.g. ref. \cite{Sarazin1988}), and the full hydrostatic
pressure equilibrium or `iHSE' model.  Both models assume a
one-parameter isothermal temperature profile, and the only difference
comes in the gas density distribution.

The Beta model has gas density profile
\be
\rho_{\rm gas}(r) = \frac{\rho_{\rm gas}(0)}{\left[1+(r/r_{\rm
      c})^2\right]^{3\beta/2}} \,, 
\ee
where $\rho_{\rm gas}(0)$ is a constant relating to the density at the
centre of the cluster, $r_{\rm c}$ is the characteristic scale of the gas
density profile, and $\beta$ is a parameter that gives the shape of the
profile.  This model is often used in cluster modelling.

In the iHSE model the gas is in full hydrostatic pressure
equilibrium, with a potential defined by the NFW profile \cite{NFW}. 
Solving for this potential, the gas profile is
found to be
\be
\rho_{\rm gas} = \rho_{\rm gas}(0)\exp\left[-\frac{4\pi G \rho_{\rm s}
   r_{{\rm s}}^{\prime 2}
  \mu}{kT}\left(1-\frac{\log(1+r/r_{\rm s})}{r/r_{\rm
      s}^{\prime}}\right)\right]\,,  
\ee
where $G$ is Newton's gravitational constant, $\mu$ is the mean mass
per particle (here set to be 0.59), $k$ is the Boltzmann constant, and
$T$ is the temperature. Finally $r_{\rm s}^{\prime}$ is the
characteristic scale of the dark matter halo, as defined in the NFW
profile. Note that the characteristic scale of the gas density in the
Beta model ($r_{\rm s}$) and the characteristic scale of the dark
matter density in the iHSE model ($r_{\rm s}^{\prime}$) are very
different parameters and cannot really be directly compared.

They generated simulated data assuming contemporary facilities:
observations of low-redshift clusters in the optical with MegaCam at
CFHT, and at 30 GHz with the extended VSA \cite{Lancaster:2004in}. 
They then went on to consider next-generation experiments,
with mock lensing observations on the ESO Wide Field Imager and mock
SZ data from AMI \cite{Zwart2008}. They used MCMC methods to fit
the SZ data for $M_{\rm gas}$ and the weak gravitational lensing data
for $M_{200}$, assuming uniform priors on both of these
parameters. The ratio of these two gives the gas fraction of the
cluster $f_{\rm gas}h$.

\begin{figure}
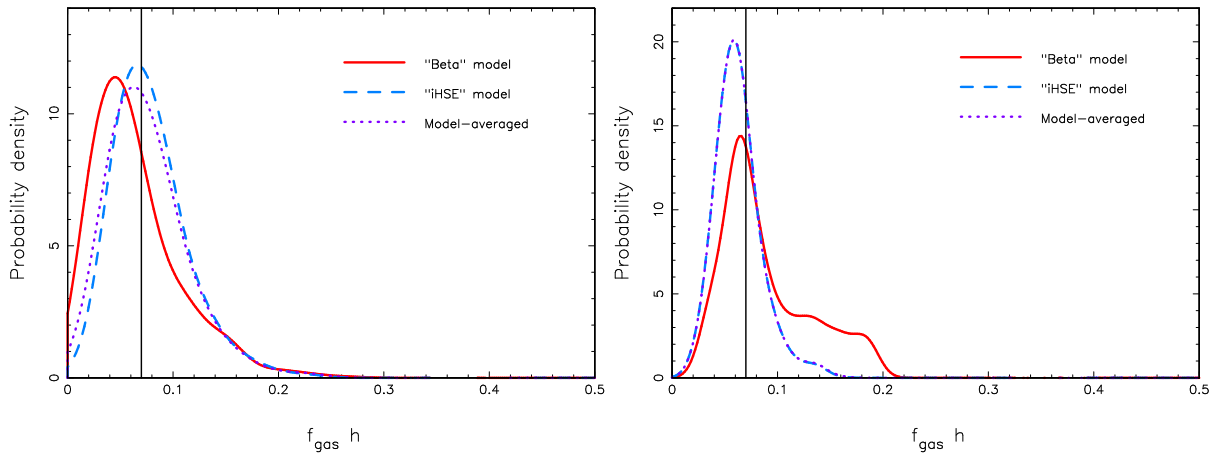

\centering
\includegraphics[scale=0.5]{current_iHSE_001_fgas.ps}
\includegraphics[scale=0.5]{nextgen_iHSE_001_fgas.ps}
\caption{Marginalized probability distributions of cluster
  cosmological parameters from mock current (top) and future (bottom)
  experiments. The true cluster model is iHSE. $P(f_{{\rm
      gas}}h|D,{\rm Beta})$ and $P(f_{{\rm gas}}h|D,{\rm iHSE})$ are
  plotted with full and dashed lines respectively; the dotted curve
  shows the result of model averaging, $P(f_{{\rm gas}}h|D)$. In the
  next-generation case, the model-averaged curve is indistinguishable
  from the true model curve, as the evidence ratio between the two
  models is large. (From ref. \cite{Marshall2003})}
\label{fig:bma_fgas}
\end{figure}

The posterior distributions for the gas fraction parameter are shown
in Figure \ref{fig:bma_fgas}. Using the current data, the evidence
values of the two models were equal within the numerical errors, and
so the model-averaged posterior is simply the average of the two
posteriors. In contrast, when using next-generation data the evidence
of the true model was much larger than that the incorrect model
(though the precise numerical values depended on what was assumed
about the primordial CMB contamination), and so the model-averaged
posterior was indistinguishable from the true model posterior.

\subsection{Distance measurements}

The measurement of distances is another area in astrophysics where
Bayesian methods are useful. Cepheid variable stars are established as
distance indicators, but calibrating the Barnes--Evans Cepheid surface
brightness relation \cite{BarnesEvans1976} of these objects is still a
challenge. Only relatively recently have interferometric measurements
of their angular diameter allowed them to be calibrated geometrically,
almost fully independent of all other astronomical distance scales \cite{Nordgren2002}.

Barnes et al. \cite{Barnes:2003ar} applied Bayesian methods to this problem. They
created an empirical model for the variation in magnitude with time
(known as the light curve) and radial velocity with time, fitting them
to Fourier polynomials of unknown order (here $M$ is the order of the
velocity model and $N$ is the order of the magnitude model). Rather
than fix the values of $M$ and $N$, they used Reversible Jump MCMC and
let them vary. Each polynomial has amplitude and phase, and there is
also a constant term, so the total number of parameters is $2M+1$ for
the velocity data and $2N+1$ for the magnitude data. They also fit for
the mean angular diameter, some hyperparameters relating to accuracy,
an unknown phase shift between the magnitude and velocity variation
and (most importantly) the distance. They put a uniform prior on the
order of the Fourier polynomials, with a cut-off value.

They used an ensemble of thirteen Cepheids selected due to their
high-quality photometry and availability of radial velocity data. They
found for most of the Cepheids that more than one order had
significant posterior probability (i.e.\ more than one value of $M$ or
$N$ had significant evidence associated with it). They compared their
results to the traditional method, where the order was selected `by
eye', usually with the criterion that the polynomial is terminated when
the scatter in the polynomial approaches the uncertainty in the
data. They found excellent agreement between the results from their
method and those of the previous more traditional methodology.

\subsection{Star Classification}

The classification of objects has been a common application of
Bayesian methods in astrophysics, where a model for the type of object
can be selected using the evidence (for example see refs. \cite{HM} and \cite{SO}). 
Since the variables involved
in the models are often very different, a model-averaging procedure
does not necessarily lend itself to such problems. However, there are
cases where the outputs of the different classification models can be
combined in such a manner.

In ref. \cite{Debosscher2007} the authors were interested in the fast
classification of variable stars using neural networks. The intention
is that this machine-learning approach can be applied to the large
numbers of objects that will be detected by the CoRoT, Kepler, and
Gaia satellite missions in an automated fashion. They use a
multi-layer perceptron, but instead of using the maximum likelihood
estimate from a single network, they use Bayesian Model Averaging to
combine the output from a number of different networks, effectively
using the networks as predictive models. They tested the Bayesian
Averaging Artificial Neural Network approach against two other
approaches, the $k$-dependent Bayesian Classifier \cite{Sahami} and
the Support Vector Machines classifier \cite{Gunn1997}. They found
that none of the methods particularly outperformed the others, all
giving rather disappointing misclassification rates. The Bayesian
Averaging Artificial Neural Network proved to be the exception in one
case, that of the eclipsing binaries, achieving a correct
classification rate of 99.73\%.

This use of Bayesian Model Averaging is very interesting, as it
illustrates how it can be applied in a situation where the underlying
physical model is not known or well understood. In this case there is
almost no physical model; the Artificial Neural Networks predict the
brightness of the transients through empirical learning. Hence there
is no prior belief about which model must be better than the other.

\section{Conclusions}
\label{sec:Conc}

In this review we have summarised the application of Bayesian Model
Averaging in a number of areas of astrophysics, namely cosmological
model averaging for parameters associated with inflation and the dark
energy, measurements of the baryonic gas fraction in galaxy clusters,
calibrating accurate distances from Cepheid variable stars, and
automated star classification using neural networks.  We found that
these applications fell broadly into one of two types, with model
averaging being applied either in the case where the
underlying physical model was uncertain, or the situation
where some details about the data acquisition or modelling was
uncertain. In all of these cases the model likelihood (evidence) was
not able to discriminate between the different models, and so they all
made some contribution to the averaged posterior.

There are still a number of research areas in astrophysics where
Bayesian model averaging has not been implemented, even though
Bayesian methods and model selection have been adopted. It is common
for data to be unable to select a single preferred model; quoting
parameter uncertainties based on selecting one particular model may
then misrepresent the actual uncertainty, and can be directly
addressed by Bayesian Model Averaging. Note that including model
uncertainty does not necessarily increase the uncertainty on model
parameters, e.g.\ tighter constraints are obtained on the dark energy
equation of state if one includes the cosmological constant model,
given its high evidence and fixed value of $w = -1$.

Areas we feel would benefit from the model-averaging approach include
questions where the parameters of some signal have to be estimated in
the presence of some unknown number of contaminating components. This
would include, for example, gravitational wave astronomy and
sub-millimetre and radio imaging surveys. Then there are cases where
some data products are estimated with some model assumed, even though
the evidence does not decisively select that model against other
candidates, for instance orbital properties of extrasolar planets in
systems where additional planets are marginally detected.

In Bayesian model averaging (as with any Bayesian method) the results
have some dependence on the choice of model priors, especially as in
the cases were model averaging is useful the model likelihoods will be
similar. This should not deter the reader from applying such methods,
as they simply require them to consider more deeply which models they
consider more reasonable or likely before examining the data.
Ultimately model averaging encodes and combines uncertainty at both
the model and parameter level, and so provides a more complete
description of the current state of knowledge.

\section*{Acknowledgments}

We thank the referees for their helpful comments and suggestions.
D.P.\ was supported by the Australian Research Council through a
Discovery Project grant, and A.R.L.\ by the Science and Technology
Facilities Council [grant number ST/I000976/1]. A.R.L.\ thanks the
Institute for Astronomy, University of Hawai`i, for hospitality while
this work was carried out.

\label{lastpage}

\end{document}